Research Paper/

# Semi-Analytical Modeling of Transient Stream Drawdown and Depletion in Response to Aquifer Pumping

Bwalya Malama

Corresponding author: Natural Resources Management and Environmental Sciences, California Polytechnic State University, San Luis Obispo, CA 93407; 805 756-2971; fax: 805 756-1402; bmalama@calpoly.edu

Ying-Fan Lin

Department of Civil Engineering, Chung Yuan Christian University, Taoyuan, Taiwan; yflin1110@cycu.edu.tw

Kristopher L. Kuhlman

Nuclear Waste Disposal Research & Analysis Department, Sandia National Laboratories, Albuquerque, NM; klkuhlm@sandia.gov

**Article Impact Statement:** *Streams have finite capacity to store water and undergo transient drawdown and depletion when a hydraulically connected aquifer is pumped.*

## Abstract

Analytical and semi-analytical models for stream depletion with transient stream stage drawdown induced by groundwater pumping are developed to address a deficiency in existing models, namely, the use of a fixed stream stage condition at the stream-aquifer interface. Field data are presented to demonstrate that stream stage drawdown does indeed occur in response to groundwater pumping near aquifer connected streams. A model that predicts stream depletion with transient stream drawdown is developed, based on stream channel mass conservation and finite stream channel storage. The resulting models are shown to reduce to existing fixed-stage models in the limit as stream channel storage becomes infinitely large, and to the confined aquifer flow with a no-flow boundary at the streambed in the limit as stream storage becomes vanishingly small. The model is applied to field measurements of aquifer and stream drawdown, giving estimates of aquifer hydraulic parameters, streambed conductance and a measure of stream channel storage. The results of the modeling and data analysis presented herein have implications for sustainable groundwater management.

## Introduction

Understanding and quantitatively predicting stream-aquifer interactions is critical to manage both these systems, while demand for water is increasing. In 2014, the state of California passed the Sustainable Groundwater Management Act (SGMA) in which stream depletion is listed among the six metrics for determining the sustainability of its groundwater basins (Owen et al., 2019). This is in recognition that groundwater pumping from aquifers overlain or bounded by streams often leads to reduced stream flows, with undesirable impacts on both human use and ecosystem function (Winter et al., 1998; Bowen et al., 2007; Yu and Chu, 2010; Foglia et al., 2013; Zipper et al., 2018; Tolley et al., 2019; Kwon et al., 2020) ultimately leading to dry streambeds or disconnected stream-groundwater systems. Theis (1941) was among the first to develop a model for stream depletion arising from groundwater pumping from a confined aquifer, with depletion defined as the decrease in stream discharge. Theis (1941) used the earlier model of Theis (1935) developed for a laterally infinite aquifer, together with the principle of linear superposition, to simulate a constant-head boundary condition at the stream-aquifer interface. Glover and Balmer (1954) extended upon this work with a closed-form function of the model, later tabulated by Jenkins (1968) who also introduced the Stream Depletion Factor concept.

Hantush (1965) made the next notable advancement by introducing a semi-pervious streambed with a general (Robin) boundary condition at the stream-aquifer interface, where the flux across the streambed was treated as proportional to the differential head across the streambed of known thickness, with stream stage held constant. Intaraprasong and Zhan (2009) expanded upon this work, solving the transient groundwater flow equation throughout the streambed with stream stage prescribed as a time-dependent function. Chan (1976) and Asadi-Aghbolaghi and Seyyedian (2010) generalized application of the superposition principle to confined aquifer flow domains bounded laterally by intersecting streams. According to Zlotnik et al. (1999), Grigoryev (1957) and Bochever (1966) were the first to consider a partially penetrating stream under steady state (Thiem-type) flow conditions in an infinite domain. Others have extended these analytical stream depletion models to cases of partially penetrating streams (Zlotnik and Huang, 1999; Zlotnik et al., 1999; Butler et al., 2001; Fox et al., 2002; Zlotnik, 2004; Butler et al., 2007; Zlotnik and Tartakovsky, 2008).

All the analytical and semi-analytical models referenced above predict confined aquifer drawdown as monotonically increasing with time before attaining a steady state. Hunt (2009), however, noted that numerous field observations (Hunt et al., 2001; Fox et al., 2002; Nyholm et al., 2002; Kollet and Zlotnik, 2003; Lough and Hunt, 2006) have shown that confined aquifer drawdown curves near a stream undergoing depletion mimic unconfined aquifer drawdown curves of delayed yield and without a permanent steady-state regime. Hunt (2009) attributed this behavior to leakage into a confined aquifer from an overlying unconfined aquifer. Stream depletion models for cases where the aquifer is unconfined have thus been developed by Hunt, et al. (2001) and Hunt (2003, 2008, 2009), who also used a fixed-stage condition with a Robin boundary condition or sink/source function in the governing flow equation for stream-aquifer fluid exchange. None of the confined aquifer fixed-stage stream depletion models reviewed above can predict the "delayed yield" behavior alluded to by Hunt (2009) hence the adoption of unconfined aquifer flow. A detailed review of the literature on different configurations of pumping-induced stream depletion problems has been provided by Huang et al. (2018), where the Dirichlet and Robin boundary conditions are identified as the only ones used for such problems.

Numerical models, such as MODFLOW (Harbaugh, 2005) and MIKE SHE (Refsgaard et al., 2010), through their respective stream packages (e.g., SFR1 and SFR2 in MODFLOW), treat stream-aquifer mass exchange in a similar manner to the analytical and semi-analytical models discussed above by treating flow across the streambed as directly proportional to the difference between stream and aquifer head (Prudic, 1989; Prudic, et al., 2004). However, MODFLOW's current stream package, SFR2, allows for the specification of spatially and temporally variable stream stage as dependent of stream discharge through stream hydrographs (Prudic et al., 2004; Harbaugh, 2005; Niswonger and Prudic, 2005) or nonlinear empirical formulas such as the Manning equation (Prudic et al., 2004), allowing for the effect of time-dependent stream stage to be considered. The use of the Manning equation, which relates stream stage to stream discharge and is nonlinear, requires iterative methods at every time step during computations to determine stream discharge and stage response to pumping. Current versions of MODFLOW (Panday et al. 2013; Langevin et al. 2017) include an approximation to the Richards equation to simulate unsaturated flow beneath streams when stream drawdown exceeds stream depth. The numerical approach, because of the nonlinearity of the flow problem, can be computationally demanding.

Given the limitations of the semi-analytical stream depletion models and the computational demands of numerical models, an alternative modeling approach is proposed here where a new boundary condition is introduced and imposed at the stream-aquifer interface by invoking the mass-balance principle and introducing the concept of finite stream channel storage. Two semi-analytical models are developed for the cases of non- or minimally-penetrating streams (NPS) and fully-penetrating streams (FPS) in a confined aquifer, taking into account the effect of finite stream channel storage. The solutions are validated by comparing them with a numerical model based on the finite-element method (FEM) and with field observations of aquifer and stream drawdown. The details of the validation exercise are included in the Supporting Information. While the approach used here does not currently include stream routing, it is a significant improvement over existing analytical and semi-analytical models and predicts the delayed yield behavior discussed in Hunt (2009). The NPS model is applied to field observations of stream and aquifer drawdown in a parameter estimation exercise by fitting the models to both aquifer and stream drawdown data to demonstrate the practical application of the approach.

## Methods

In the following, we describe the mathematical formulation of the new stream-aquifer interaction term, develop semi-analytical solutions for the two cases already mentioned above, and apply the model to field observations of stream and aquifer drawdown. We consider the case of groundwater flow to a fully penetrating pumping well completed in a homogeneous confined aquifer near a hydraulically connected stream. For the stream, we consider two cases, namely (a) a non-penetrating stream (NPS), where the stream flows atop the aquifer and (b) a fully penetrating stream (FPS), where the stream cuts across the entire thickness of the aquifer. For the FPS case, we distinguish between the case in which flow in the aquifer occurs on both sides of the stream, which we term FPS case 1 (FPS-1), and one in which flow in the aquifer occurs only in the half-space containing the pumping well, which we term FPS case 2 (FPS-2). The conceptual models of the problem described here are shown schematically in Figure 1. To solve the flow problem described below, we make the following assumptions:

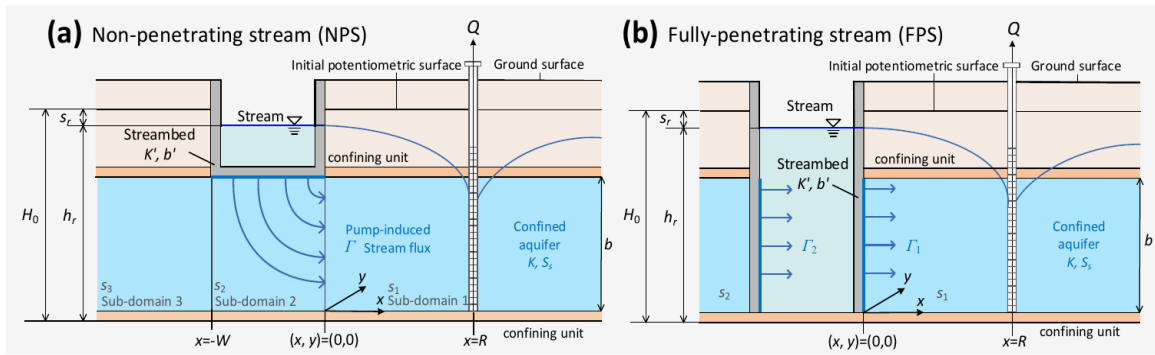

Figure 1. Conceptual model of the stream-aquifer system used for the (a) non-penetrating stream (NPS) and (b) fully penetrating stream (FPS) models derived herein. For the NPS case, the depicted stream stage is the averaged value across the stream channel width.

1. The aquifer is fully confined and homogeneous but anisotropic with horizontal hydraulic conductivities $K_x$ and $K_y$ [L/T] in the x- and y-directions, respectively, has uniform thickness $b$ [L], and is of infinite lateral extent away from the stream.
2. Groundwater is pumped at a constant volumetric flow rate, $Q$ [L$^3$/T], from a line sink located at a distance R [L] from the stream bank.
3. The aquifer interacts with the stream across a streambed with uniform hydraulic conductivity $K'$ [L/T], thickness $b'$ [L], and conductance $\beta = K'/b'$ [1/T], as in Hantush (1965).
4. The effects of streambed storage are neglected, with the focus being only on the effects of stream channel storage (Huang *et al.*, 2020; Xiong et al., 2021).
5. Stream stage, $h_r(y,t)$ [L], responds to groundwater pumping and experiences a transient drawdown response. It is measured relative to the bottom of the aquifer with stream drawdown defined as $s_r(y,t) = H_0 - h_r(y,t)$ [L], where $H_0$ [L] is the initial stream stage, which is set equal to the initial aquifer hydraulic head; the stream-aquifer system is assumed to be initially at equilibrium.

These assumptions are adopted to simplify a complex reality and improve the tractability of the mathematical problem. In the following, we describe the flow equation and the associated initial and boundary conditions required to solve the flow problem.

## Governing Equations of Flow

The governing equation of the two-dimensional confined aquifer flow problem considered in this work is (Fox et al., 2002)

$$S_s \frac{\partial s}{\partial t} = K_x \frac{\partial^2 s}{\partial x^2} + K_y \frac{\partial^2 s}{\partial y^2} + f_s(x, y, t) \quad (1)$$

where $s = H_0 - h(x, y, t)$ [L] is aquifer drawdown, $(x, y)$ [L] are spatial coordinates in the horizontal plane, $t$ [T] is the elapsed time from the onset of pumping, $S_s$ [1/L] is aquifer specific storage, and $f_s(x, y, t)$ is a sink/source function [1/T]. The $x$ coordinate axis is perpendicular to the axis of the stream channel with the origin set on the stream bank closest to the pumping well. The $y$ axis origin is located at the point where the line perpendicular to the pumping well and the stream intersects the closest stream bank and extends from $y \to \infty$ to $y \to -\infty$.

The aquifer flow problem is solved subject to the usual homogeneous initial condition and far-field boundary conditions. For simplicity, the stream-aquifer flow system is assumed to be initially at equilibrium. For cases where a background uniform flow field exists, the principle of superposition may be invoked to apply the solution developed herein.

Table 1. Nomenclature

| Variable | Definitions | Variable | Definitions |
|---|---|---|---|
| $b$ | Aquifer thickness [L] | $s_{obs}$ | Observed drawdown [L] |
| $b'$ | Thickness of streambed [L] | $s_r$ | Stream drawdown [L] |

| Symbol | Description | Symbol | Description |
|---|---|---|---|
| $C_r$ | Stream channel storage coefficient [-] | $t$ | Elapsed time from start of pumping [T] |
| $f_s$ | Sink/source function [1/T] | $t_n$ | Time at *n*-th observation [T] |
| $F$ | Objective function (sum of squared residuals) | $W$ | Width of the stream [L] |
| $H_0$ | Initial stream and aquifer hydraulic head [L] | $(x, y)$ | Spatial coordinates [L] in horizontal plane |
| $h_r$ | Stream stage [L] | $Y$ | Set of model parameters |
| $K_x, K_y$ | Directional hydraulic conductivity [L/T] | $Y_{opt}$ | Set of estimated model parameters |
| $K'$ | Streambed hydraulic conductivity [L/T] | $\alpha = K_x/S_s$ | Hydraulic diffusivity [L$^2$/T] |
| $N$ | Number of temporal drawdown observations | $\beta = K'/b'$ | Streambed conductance [1/T] |
| $q_r$ | Point-wise depletion flux [L/T] | $\gamma$ | Euler's constant $\simeq 0.577216$ |
| $QQ$ | Aquifer well pumping rate [L$^3$/T] | $\Gamma$ | Stream-aquifer mass transfer function [L/T] |
| $Q_r$ | Stream depletion rate [L$^3$/T] | $\kappa = K_y/K_x$ | Horizontal anisotropy ratio [-] |
| **$Q_s$** | **Stream discharge [L$^3$/T]** | | |
| $R$ | Pumping well [L] distance from stream bank | $\Delta A_r$ | Unit area of streambed [L$^2$] |
| $S_s$ | Specific (elastic) aquifer storage [1/L] | $\Delta h_r$ | Unit change in stream stage [L] |
| s | Aquifer drawdown [L] | $\Delta V_w$ | Unit volume of stream water [L$^3$] |
| | | $\Delta y$ | Unit distance [L] in *y*-direction |

**Non-Penetrating Stream**

We first consider the case of a non-penetrating stream (NPS), where the stream flows atop the aquifer to simulate the case where the stream has incised through the upper confining unit and minimally penetrates the aquifer. The schematic of the conceptual model of this case is shown in Figure 1(a). To solve the flow problem, aquifer drawdown is defined in a piecewise manner (see supplemental materials) with the sink/source function, $f_s(x, y, t)$, defined as

$$f_s(x,y,t) = \begin{cases} \frac{1}{b}Q\delta(x-R)\delta(y) & x > 0 \\ \Gamma/b & -W \leq x \leq 0 \\ 0 & x < -W \end{cases} \quad (5)$$

where $\delta(.)$ is the Dirac delta function [1/L], $W$ is stream width [L], and $\Gamma$ [L/T] is the stream-aquifer mass-transfer function through the base of the stream. The term $Q\delta(x-R)\delta(y)$ represents the pumping well line sink function at $(R, 0)$.

The stream-aquifer mass-transfer function, $\Gamma$, at the streambed is

$$\Gamma = \beta[s(x,y,t) - s_r(y,t)] \quad -W \leq x \leq 0. \quad (6)$$

Additionally, continuity conditions for drawdown and flux at $x = 0$ and $x = -W$ are enforced at the two stream banks (see supplemental materials for details).

It should be noted that for the NPS case vertical flow across the streambed is treated as a sink/source term in equation (5) defined by the mass-transfer function in equation (6), where stream stage and the associated drawdown, $s_r(y,t)$, are functions of both $y$ and time $t$. This dependence on time is what distinguishes the present work from fixed-stage models where stage is held constant at $h_r = H_0$ and stream drawdown vanishes identically ($s_r \equiv 0$). To allow for a temporally variable stream stage and drawdown, an additional condition is required in the stream as discussed subsequently.

**Fully Penetrating Stream**

In the second case the stream has fully incised through both the thicknesses of the upper confining unit and the aquifer, commonly referred to as a fully penetrating stream (FPS). A schematic of the conceptual model for this case is shown in Figure 1(b). Aquifer drawdown again is defined in a piecewise manner with $s_1(x, y, t)$ being the drawdown of the aquifer in the half-space with the pumping well, and $s_2(x, y, t)$ is the drawdown in the far side half-space. We refer to the FPS case where aquifer flow occurs on both sides of the stream as FPS case 1 (or FPS-1). For the case where the flow on the far-side $(x < -W)$ half-space is neglected, only the drawdown on the pumped half-space is considered. Such a case may occur when the stream flows along a fault line where the far-side bank comprises an impermeable formation. We refer to this as FPS case 2 (or FPS-2).

The boundary condition imposed at the stream-aquifer interface is specified as

$$-K_x \frac{\partial s_1}{\partial x}\bigg|_{x=0} = \Gamma_1 \quad (12)$$

for the pumped half-space, and

$$-K_x \frac{\partial s_2}{\partial x}\bigg|_{x=-W} = \Gamma_2 \quad (13)$$

on the far side, where the mass-transfer functions $\Gamma_1$ and $\Gamma_2$ are defined as

$$\Gamma_1 = -\beta[s_r(y, t) - s_1(x, y, t)|_{x=0}] \quad (14)$$

and

$$\Gamma_2 = -\beta[s_r(y, t) - s_2(x, y, t)|_{x=-W}] \quad (15)$$

for the pumped-side and far-side stream-aquifer interfaces, respectively. Here, we assume that the two interfaces have the same conductance, $\beta$. For the FPS-2 case, $\Gamma_2 \equiv 0$. Stream drawdown, $s_r(y,t)$, is again treated as an unknown time-dependent function, which necessitates an additional condition discussed in the following section.

**Accounting for Stream Drawdown and Channel Storage**

The flow problem is incomplete without specifying an additional condition at the stream-aquifer interface, namely a mass-balance condition applied to the stream channel. This mass balance condition states that the rate of change of mass within the stream channel equals the rate of mass transfer across the streambed induced by pumping. For simplicity, in the following development of the semi-analytical solutions, we neglect stream flow velocity effects.

For the NPS model, the mass conservation condition can be mathematically stated as

$$\Gamma = C_r \frac{\partial s_r}{\partial t} \qquad (16)$$

whereas for the FPS model

$$\Gamma_1 + \Gamma_2 = C_r \frac{\partial s_r}{\partial t} \qquad (17)$$

where $C_r$ [-] is a stream channel storage coefficient. It is defined here as a dimensionless measure of the volume of water, $\Delta V_w$, which flows through a unit area of the streambed, $\Delta A_r$, per unit change in stream stage, $\Delta h_r$ (i.e., $C_r = \Delta V_w/(\Delta A_r \Delta h_r)$). It is a measure of the volume contribution of water stored in the stream channel to aquifer flow, and is distinguished here from streambed elastic storage, which is neglected in the present study (Huang *et al.*, 2020; Xiong et al., 2021). According to Huang et al. (2020) and Xiong et al. (2021) the effect of streambed specific storage, $S'_s$, may be neglected when the condition $S'_s < (0.1R/b')S_s$ is satisfied (Huang *et al.*, 2020; Xiong, et al., 2021), where $S_s$ is aquifer specific storage, $R$ is the distance from the stream to the pumping well and $b'$ is streambed thickness. In typical physical systems, $R \gg b'$, and if streambed sediment of similar composition as the aquifer material such that $S'_s$ is of the same order of magnitude as $S_s$, the above inequality is expected to hold. We limit the present development to cases where this condition is satisfied.

For a stream channel with an idealized uniform geometric cross-sectional structure, it is possible to provide simple expressions for this parameter. For example, in the FPS case, the stream channel has a rectangular cross section, with stream width $W$ and aquifer thickness $b$, $\Delta V_w = W \Delta y \Delta h_r$, $\Delta A_r = 2b \Delta y$, leading to $C_r = W/(2b)$ if mass exchange is limited to the stream bank. For the NPS case, where a similar simple geometric profile may be adopted for the cross section of the stream channel, it can be shown that $C_r = 1.0$. Hence, for real systems where the uniform geometric profile assumption may be adopted, the stream channel storage coefficient is expected to be of the order of unity. For more complex stream channel cross-sectional geometries that vary spatially with $y$, the parameter $C_r$ can be empirically estimated by inversion of stream and aquifer drawdown data, as is demonstrate in this work. It is also possible, in principle, to develop empirical functions relating $C_r$ to the dimensionless ratio $\langle W \rangle/b'$ of the form $C_r = f(\langle W \rangle/b')$, where $\langle W \rangle$ is a spatial average of stream width along the length of the stream. For mathematical tractability, we restrict the development of the semi-analytical solutions to cases where the stream channel storage coefficient, $C_r$, is a constant. This simplification linearizes the boundary condition at the stream-aquifer interface.

Table 2. Definitions of dimensionless variables and parameters based on a characteristic length $L_c = R$, time $T_c = R^2/\alpha_x$ with $\alpha_x = K_x/S_s$, head $H_c = Q/(2bK_x)$, and flux $q_c = H_c K_x/R$.

| Symbols | Definitions | Symbols | Definitions |
|---|---|---|---|
| $s_{D,i}$ | $s_i/H_c$ | $\kappa'$ | $K'/K_x$ |
| $s_{D,r}$ | $s_r/H_c$ | $\beta_D$ | $\beta R/K_x$ |
| $x_D$ | $x/R$ | $\beta_D^*$ | $\beta_D/C_{D,r}$ |
| $y_D$ | $y/R$ | $C_{D,r}$ | $b_D C_r/S$ |
| $t_D$ | $t/T_c$ | $W_D$ | $W/R$ |
| $b_D$ | $b/R$ | $q_D$ | $q/q_c$ |
| $b_D'$ | $b'/R$ | $Q_{D,r}$ | $Q_r/Q$ |
| $\kappa$ | $K_y/K_x$ | | |

## Analytical Solutions of the Flow Problem

To solve the flow problem described above, the governing equation is first transformed into dimensionless form. The dimensionless variables and parameters that appear in the solution are defined in Table 2. In the development of the dimensionless forms, we set the characteristic length, $L_c$, as equal to the distance, $R$, of the pumping well from the nearest stream bank, $L_c = R$, which is the controlling distance for pumping induced hydraulic stress on the stream. The characteristic time is set to $T_c = R^2/\alpha_x$, where $\alpha_x = K_x/S_s$ is x-direction aquifer hydraulic diffusivity. The characteristic head and flux are then $H_c = Q/(2bK_x)$ and $q_c = H_c K_x/R$, respectively, to limit the number of free parameters in the resulting flow equations. Laplace and Fourier-cosine transforms are applied to the dimensionless governing equations, which are then solved by standard methods for ordinary differential equations. The respective inversion formulae of the transforms are finally used to numerically obtain the applicable flow solutions in space-time. The transform and inversion formulae can be found in standard textbooks of Engineering Mathematics, and interested readers may refer to reference texts (Haberman, 2012). Similar solution approaches have been used in hydrogeology literature (Butler et al., 2001).

**Case of a Non-Penetrating Stream**

The exact solution for aquifer drawdown, in transform space, for the NPS is (see Supporting Information for details of derivation)

$$\tilde{\bar{s}}_{D,NPS} = \frac{2e^{-\eta}}{p\Delta_1} \begin{cases} e^{\eta(1-x_D)} \tilde{\bar{g}}_1(p, \xi, 1.0) & x_D > 1 \\ \tilde{\bar{g}}_1(p, \xi, x_D) & 0 \leq x_D \leq 1 \\ \hat{\eta} \cosh[\hat{\eta}(x_D + W_D)] + \eta \sinh[\hat{\eta}(x_D + W_D)] & -W_D < x_D < 0 \\ \hat{\eta} e^{\eta(W_D + x_D)} & x_D \leq -W_D \end{cases} \quad (18)$$

where $\tilde{\bar{s}}_{D,NPS}$ is the Laplace (overbar) and Fourier cosine (tilde) transform of $s_D$ for the NPS case, $p$ is the dimensionless Laplace transform variable, $\xi$ is the dimensionless Fourier cosine transform variable, and with

$$\eta = \sqrt{p + \kappa \xi^2} \quad (18a)$$

$$\hat{\eta} = \sqrt{\eta^2 + \zeta} \qquad (18b)$$

$$\zeta = \frac{p\beta_D}{p + \beta_D^*} \qquad (18c)$$

$$\beta_D^* = \beta_D/C_{D,r} \qquad (18d)$$

$$\Delta_1 = 2\hat{\eta}\eta \cosh(\hat{\eta}W_D) + (\eta^2 + \hat{\eta}^2)\sinh(\hat{\eta}W_D) \qquad (18e)$$

$$\tilde{\bar{g}}_1(p,\xi,x_D) = \hat{\eta}e^{\eta x_D}\cosh(\hat{\eta}W_D) + \tilde{\bar{g}}_2(p,\xi,x_D)\sinh(\hat{\eta}W_D) \qquad (18f)$$

$$\tilde{\bar{g}}_2(p,\xi,x_D) = \eta\cosh(\eta x_D) + (\hat{\eta}^2/\eta)\sinh(\eta x_D) \qquad (18g)$$

Here, $\beta_D = R\beta/K_x = \kappa'/b_D'$ is the dimensionless stream conductance, $W_D = W/R$ is the dimensionless stream channel width, and $C_{D,r} = b_D(C_r/S)$ is the dimensionless ratio of the stream channel storage coefficient to aquifer storativity, $S = bS_s$, scaled by the normalized aquifer thickness.

The corresponding solution for stream drawdown in transform space is given by

$$\tilde{\bar{s}}_{D,r} = \frac{\tilde{\bar{s}}_D(p,\xi,x_D)}{1 + p/\beta_D^*}, \quad -W_D < x_D < 0, \qquad (19)$$

where $\tilde{\bar{s}}_{D,r}$ is the Laplace and Fourier cosine transform of dimensionless stream drawdown $s_{D,r}$. Upon inversion from transform space, Equation (19) may be used to compute stream drawdown induced by pumping from a well completed in a confined aquifer. Space-time stream and aquifer drawdown are obtained by numerical inversion of the Fourier cosine and Laplace transforms using numerical quadrature and the Stehfest (1970) algorithm implemented using Mathematica and MATLAB. Mathematica and MATLAB algorithms of numerical integral and inversion are used to evaluate the semi-analytical solutions presented above. The Mathematica and MATLAB scripts developed by the authors for this purpose can be found in the hyperlinks provided with the Supplemental Information.

**Case of a Fully Penetrating Stream**

As detailed in the Supporting Information, the aquifer drawdown solution for the FPS-1 model, allowing for flow on both sides of the stream, is given by

$$\tilde{\bar{s}}_{D,\text{FPS1}}(p,\xi,x_D) = \frac{2e^{-\eta}}{p\Delta_2} \begin{cases} e^{-\eta(x_D-1)}\tilde{\bar{g}}_3(p,\xi,1.0) & x_D > 1 \\ \tilde{\bar{g}}_3(p,\xi,x_D) & 0 \leq x_D < 1 \\ p\beta_D\beta_D^* e^{\eta(W_D+x_D)} & x_D \leq -W_D \end{cases} \quad (20)$$

where $\tilde{\bar{s}}_{D,\text{FPS1}}$ is the Laplace and Fourier cosine transform of aquifer drawdown, $s_D$, for the FPS case with flow on both sides of the stream and

$$\tilde{\bar{g}}_3(p,\xi,x_D) = \chi_1 \cosh(\eta x_D) + \chi_2 \sinh(\eta x_D) \quad (20a)$$

$$\chi_1 = \frac{\Delta_2}{\beta_D + \eta} + p\beta_D\beta_D^* \quad (20b)$$

$$\chi_2 = \frac{\beta_D \Delta_2}{\eta(\beta_D + \eta)} - p\beta_D\beta_D^* \quad (20c)$$

$$\Delta_2 = p(\beta_D + \eta)[p(\beta_D + \eta) + 2\eta\beta_D^*]. \quad (20d)$$

The corresponding stream drawdown solution in transform space is

$$\tilde{\bar{s}}_{D,r,\text{FPS1}}(p,\xi) = \frac{2}{\Delta_2}(\eta + \beta_D)\beta_D^* e^{-\eta}. \quad (21)$$

The space-time form of the solution is obtained by numerical inversion of the transforms as stated previously.

For the case where there is no flow across the far-side stream-aquifer interface, with $\Gamma_2 \equiv 0$, the solution reduces to

$$\tilde{\bar{s}}_{D,\text{FPS2}}(p,\xi,x_D) = \frac{2}{p\eta\Delta_3}\begin{cases} g_5(p,\xi,1.0)e^{-\eta x_D} & x_D \geq 1 \\ g_5(p,\xi,x_D)e^{-\eta} & 0 \leq x_D < 1 \end{cases} \quad (22)$$

where $\tilde{\bar{s}}_{D,\text{FPS2}}$ is the Laplace and Fourier-cosine transform of $s_D$ for the FPS case with flow only on the well side of the stream, $\Delta_3 = \eta(p + \beta_D^*) + p\beta_D$, and

$$g_5(p,\xi,x_D) = \eta(p + \beta_D^*)\cosh(\eta x_D) + p\beta_D \sinh(\eta x_D) \quad (23)$$

The full derivation of this solution is included in the Supporting Information for completeness. The corresponding solution for dimensionless stream drawdown is

$$\tilde{\bar{s}}_{D,r,\text{FPS2}}(p,\xi) = \frac{2}{p\Delta_3}\beta_D^* e^{-\eta} \quad (24)$$

which upon inversion gives dimensionless stream drawdown, $s_{D,r}(y, t_D)$ as a function of time, $t_D$, and position along the stream channel, $y_D$.

**Stream Depletion Solution**

Stream depletion, $Q_r$ $[L^3/T]$, defined as the volume rate of flow captured from the stream by a pumping well, is obtained by integrating the streambed flux along the length of the stream. The depletion flux, $q_r$ $[L/T]$, in dimensionless form, is

$$q_{D,r} = C_{D,r} \frac{\partial s_{D,r}}{\partial t_D} \quad (25)$$

where $q_{D,r} = q_r/(2bR/Q)$. In the Laplace and Fourier cosine transform domain, equation (25) becomes $\tilde{\bar{q}}_{D,r} = pC_{D,r}\tilde{\bar{s}}_{D,r}$, which upon inverting the Fourier cosine transform becomes (Povstenko, 2015)

$$\bar{q}_{D,r} = pC_{D,r} \int_0^\infty \tilde{\bar{s}}_{D,r} \cos(\xi x_D)\, d\xi. \quad (26)$$

Therefore, it follows that

$$\bar{Q}_{D,r} = \begin{cases} C_{D,r} \int_{-\infty}^{\infty} p\bar{s}_{D,r} dy_D & \text{FPS} - 1,2 \\ \dfrac{C_{D,r}}{b_D} \int_{-\infty}^{\infty} \int_{-W_D}^{0} p\bar{s}_{D,r} dx_D\, dy_D & \text{NPS} \end{cases} \quad (27)$$

where $Q_{D,r} = Q_r/Q$ is the normalized stream depletion rate (SDR), which expresses the depletion rate, $Q_r$, as a fraction of the pumping rate, $Q$. Equation (27) includes improper integrals, which can be time consuming to evaluate numerically. For practical purposes we estimate it as a definite integral over the interval $y_D \in [-L_D, L_D]$, where $L_D \approx R_{D,\infty}$, the radius of influence of the pumping well. Additional information on the radius of influence for different aquifer and pumping well configurations can be found in Bresciani et al. (2020).

## Application to Field Observations

To test the assumption of fixed stage during groundwater pumping, we monitored the response of a stream to groundwater pumping for irrigation. The null hypothesis in this case is that streams act as constant-head boundaries or as sources of the fixed-stage mass-transfer type, supplying recharge to an aquifer indefinitely during groundwater pumping. In the following, we provide evidence from field observations that a stream in hydraulic contact with an aquifer can experience a measurable transient drawdown response to groundwater pumping.

**Study Site Description**

The study site for this work is situated in the agricultural fields of the California Polytechnic State University (Cal Poly), San Luis Obispo, California. A map of the study site is shown in Figure 2. The site is in an alluvial basin underlain with a shallow confined aquifer of gravel and sand, which is underlain with low permeability metavolcanic bedrock. It is bounded above by a thin near-surface layer of variably saturated clay or clay-rich sediment of very low permeability, which constitutes the upper confining unit with an average thickness of 13 m. The aquifer has an average thickness of 11 m locally, as determined from drilling logs. A stream, namely Stenner Creek, flows across the study site atop the aquifer in a nearly northwest-to-southeast direction, incising through the entire thickness of the confining layer that overlies the aquifer. The stream is in direct hydraulic contact with the aquifer, with a streambed made up of the same sand and gravel formation as the aquifer. The stream only minimally penetrates the aquifer. During the summer low flows, the discharge rate of the stream is on the order of $Q_s \sim 5 \times 10^{-4}$ m³/s. The groundwater basin in which the aquifer is situated has been designated as having medium priority in the implementation of the Sustainable Groundwater Management Act (SGMA) enacted in 2014 in California. In SGMA, depletion of surface waters is specifically identified as one of the metrics by which the state establishes whether a groundwater basin is managed sustainably (Owen et al., 2019).

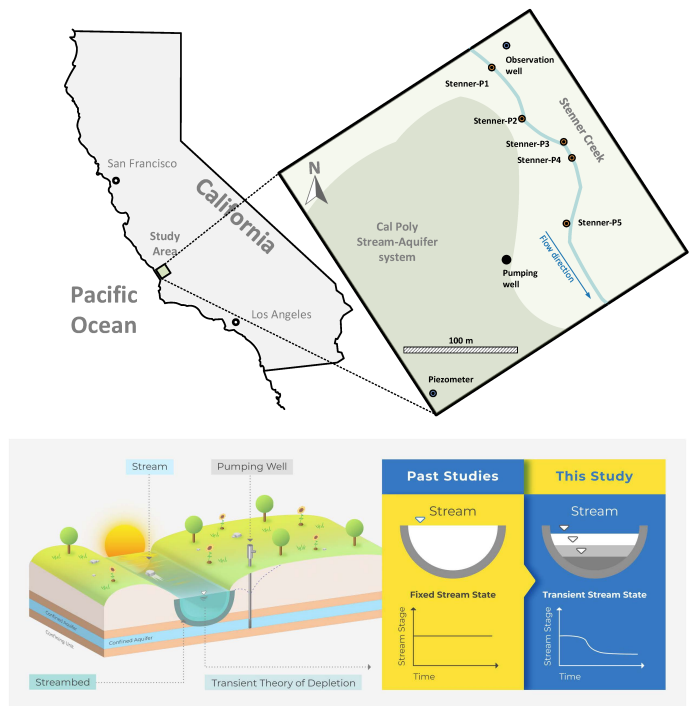

Figure 2. The map of the study site at California Polytechnic State University, San Luis Obispo, showing the observation locations along the stream and the observation well and the aquifer piezometer. A schematic of aquifer-stream flow domain at the site is included with the map.

**Monitoring of Aquifer Pumping**

The pumping well in this study has a diameter of 0.2 m and is located about 60 m southwest of the stream, as depicted in the site map in Figure 2. It is completed across the entire thickness of the aquifer and is used to pump groundwater fortnightly at a constant rate of $Q = 8.58 \times 10^{-3}$ m$^3$/s. Aquifer water levels were continuously monitored with pressure transducers in the pumping well and a nearby abandoned well located across the stream about 15 m from the far-side bank as shown on the map. Stream stage was monitored at 15-minute intervals with pressure transducers in stream channel stilling wells at the locations labeled Stenner-P1 to Stenner-P5 on the site map in Figure 2. Water-level data were collected at all these monitoring locations shown on the site map except Stenner-P4 where there was an instrument malfunction, and no data were collected. The raw water level time series data from monitoring locations are shown in Figure 3. They were collected over a period including the spring season characterized by high stream flows, and the summer when stream flows are at their lowest. Stream flow discharge rate and stage change significantly over the seasons, from the high flows associated with the end of the winter rain season to the low flows of the dry spring-summer irrigation season. This has implications for the channel storage coefficient values over the study period.

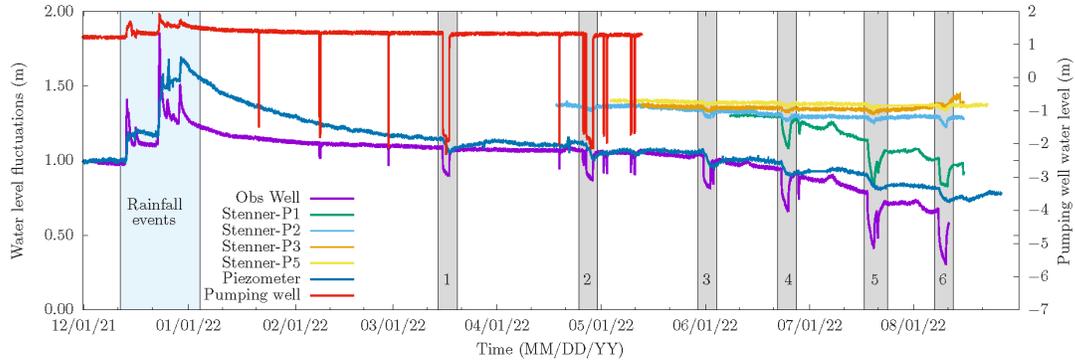

Figure 3. Time series of groundwater and stream stage fluctuations observed in the stream-aquifer system. The six pumping events are indicated in gray and marked 1-6. The scale on the left is for stream and aquifer observation wells and on the right is for pumping well water levels.

**Drawdown Analysis and Parameter Estimation**

To demonstrate the applicability of the model to field observations of aquifer and stream drawdown, a model fitting and parameter estimation exercise was conducted using observed drawdown. Drawdown data were computed by denoising and detrending the raw stream stage time series data using singular spectrum analysis and ensemble empirical mode decomposition (Faouzi and Janati, 2020; Laszuk, 2017) to remove noise from diurnal fluctuations due to riparian corridor evapotranspiration and the trend due to rainfall events (highlighted in blue in the figure) and natural discharge. Six pumping events recorded during the spring and summer irrigation season of 2022 are highlighted gray in Figure 3. The six pumping events analyzed in this work are marked in grey in Figure 3.

The stream minimally penetrates the aquifer at the study site, so the NPS solution was used to estimate aquifer hydraulic parameters, namely, hydraulic conductivity $K_x$, anisotropy ratio $\kappa$, and specific storage $S_s$, as well as the streambed conductance, $\beta$, and the stream channel storage coefficient, $C_r$. The NPS solution was coupled with the Marquardt-Levenberg optimization algorithm as implemented in MATLAB to identify parameter values that minimize the sum of squared residuals between observed and model predicted drawdown, viz.,

$$F(Y) = \sum_{n=1}^{N} [s_{\text{obs}}(t_n) - s_{\text{cal}}(t_n; Y)]^2 \qquad (28)$$

where $N$ is the total number of temporal drawdown observation time series, $s_{\text{obs}}(t_n)$, at times $t_n$, with $n = 1, 2, \ldots, N$ and $s_{\text{cal}}(t_n; Y)$ are the corresponding model computed values given the set of parameters $Y = \{K_x, S_s, \kappa, \beta, C_r\}$. The convergence criterion for the optimization process was set to $F(Y_{\text{opt}}) < 10^{-4}$, where $Y_{\text{opt}}$ is the set of parameters that minimizes the objective function defined above. The hydraulic parameters were log-transformed to constrain the optimization procedure to the positive space of the parameter values.

To provide a measure of parameter estimation uncertainty, the standard errors were computed from the diagonal elements of the covariance matrix, $\mathbf{V}$, of parameter estimates, which is computed as

$$\mathbf{V} = (J^T J)^{-1} \sigma_\epsilon^2 \qquad (29)$$

where $J$ is the relative sensitivity (Jacobian) matrix, and $\sigma_\epsilon^2$ is the variance of measurement errors, which is approximated in this work by the variance of model-observation residuals at parameter estimation optimality.

In this work, only drawdown data from the aquifer observation well and the stream stilling well Stenner-P1 were used in the analysis presented here. The pumping well coordinates were set at $(x,y) = (62, 0.0)$ m. Stenner-P1 was in the middle of the stream channel at $(x,y) = (-0.75, 183)$ m, and the aquifer observation well on the far-side of the stream at $(x,y) = (-15.0, 194)$ m. The stream channel width averaged $W = 1.5$ m during the period when the data were collected. In all the tests analyzed here, groundwater was pumped at a constant rate of $Q = 8.58 \times 10^{-3}$ m³/s to irrigate a lemon orchard for a total of 48 hours. The pumping rate increased slightly at the start of the second 24-hour irrigation period, which affected the drawdown response beyond 24-hours. Hence, only data from the first 24 hours of pumping were analyzed. Recovery phase data were not considered to simplify the analysis.

## Results

The results presented here are separated into three parts, namely (1) model predicted behavior, (2) the observed stream-aquifer drawdown at the study site, and (3) estimation of aquifer and stream hydraulic parameters from aquifer and stream drawdown.

### Model Predicted Behavior

The transient aquifer and stream drawdown behavior predicted by the model developed herein are shown in Figure 4. The Figure shows the behavior predicted by (a) the NPS model, and (b) the FPS model with the solutions FPS case 1 and 2 are marked as FPS-1 and FPS-2 on the graph.

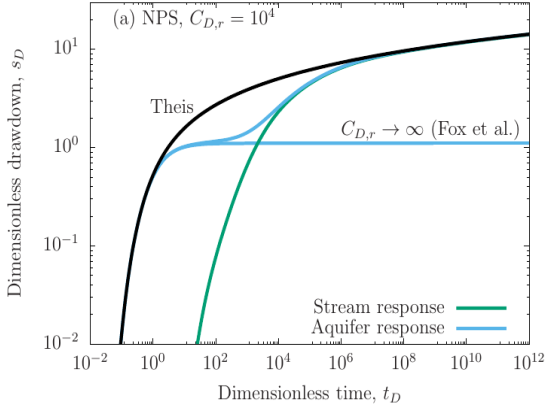

Figure 4. Transient aquifer ($s_D$) and stream ($s_{D,r}$) drawdown response predicted at $(x_D, y_D) = (0.5, 0)$ and $(x_D, y_D) = (-0.5W_D, 0)$, respectively by the (a) NPS and (b) FPS solutions for a fixed value of $\beta_D = 10$. The FPS case 1 and 2 solutions are marked as FPS-1 and F

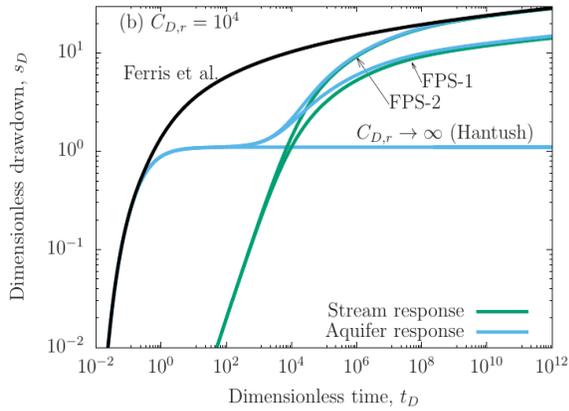

PS-2 on the graph in (b). The limiting cases of Theis (1935) and Fox et al. (2002) for the NPS solution and, of Ferris et al. (1962) (impermeable barrier), and Hantush (1965) for the two FPS solutions, are included for comparison.

The graphs show dimensionless aquifer and stream drawdown, $s_D$ or $s_{D,r}$, on the vertical axis, versus dimensionless time, $t_D$, on the horizontal axis. The aquifer drawdown curves were computed at $(x_D, y_D) = (0.5, 0)$ and the stream drawdown curves at $(x_D, y_D) = (-0.5W_D, 0)$ with a dimensionless conductance of $\beta_D = 10$. The limiting cases are included for comparison.

The models of Hantush (1965) and Fox et al. (2002) used a constant head boundary condition and correspond to the limiting case of $C_{D,r} \to \infty$ for the NPS and FPS models, respectively. The model of Theis (1935) is shown in Figure 4(a) as the limiting case of $C_{D,r} \to 0$ for the NPS solution for an impermeable barrier at the stream-aquifer interface when there is no water in the stream, while that of Ferris et al. (1962) is included in (b) as the limiting case of the FPS solutions when the stream runs dry and becomes an impermeable barrier.

The aquifer drawdown response shown in Figure 4 has a characteristic S-shape in log-time, characterized by early-, intermediate-, and late-time phases, whereas that of the stream is monotonic and delayed in time relative to the aquifer drawdown response. The early-time response follows the Theis (1935) (NPS) and Ferris et al. (1962) (FPS) solutions when water is being drawn primarily from aquifer elastic storage, while the intermediate phase follows the Fox et al. (2002) (NPS) and Hantush (1965) (FPS) solutions when the stream contribution becomes significant. The predicted late-time aquifer drawdown exceeds levels predicted by the constant-head models of Fox et al. (2002) and Hantush (1965).

The dependence of the model predicted temporal behavior of aquifer and stream drawdown on the dimensionless stream channel storage coefficient, $C_{D,r}$, is shown in Figure 5. The limiting model of Fox et al. (2002) is again included for comparison. Aquifer drawdown predicted behavior at $(x_D, y_D) = (0.5, 0)$ for the NPS solution is shown. The stream drawdown behavior was computed at the center of the stream. All the curves shown in the figure were computed for a fixed value of $\beta_D = 10$. Similar behavior is predicted for the FPS case 1 and 2 solutions, hence the graphs for these two cases are not included here for brevity.

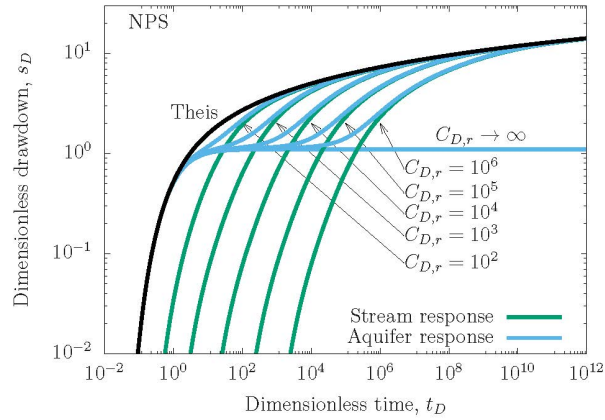

Figure 5. Model predicted temporal behavior of aquifer and stream dimensionless drawdown at $(x_D, y_D) = (0.5, 0)$ and $(x_D, y_D) = (-0.5W_D, 0)$, respectively, for different values of the dimensionless storage coefficient, $C_{D,r}$, with the dimensionless conductance fixed at $\beta_D = 10$.

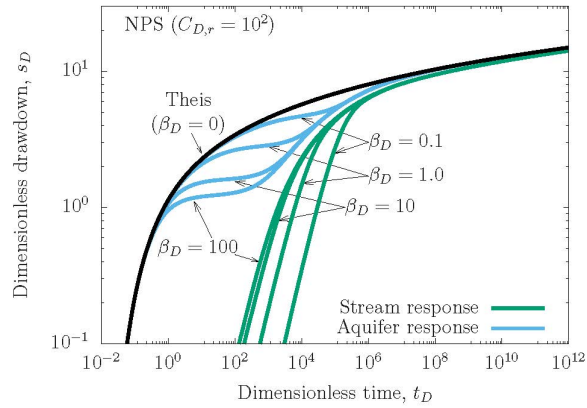

Figure 6. Model predicted temporal behavior of dimensionless aquifer and stream drawdown at $(x_D, y_D) = (0.5, 0)$ and $(x_D, y_D) = (-0.5W_D, 0)$, respectively for different values of $\beta_D$ ranging from 0 to 100.

The effect of dimensionless streambed conductance, $\beta_D$, on aquifer and stream drawdown is depicted in Figure 6 for the NPS solution. The curves in the graphs are for different values of $\beta_D$ ranging from an impermeable streambed at $\beta_D = 0$ to a highly conductive one at $\beta_D = 10$, with the dimensionless channel storage coefficient fixed at $C_{D,r} = 100$. Similar behavior is predicted by the two FPS solutions.

Drawdown derivative analysis is a useful diagnostic tool because it can improve the sensitivity of predicted model behavior to hydraulic parameters and lead to improved parameter identifiability (Chow, 1952; Bourdet *et al.*, 1983; Ferroud et al., 2018; Ferroud et al., 2019). Figure 7 shows the temporal behavior of aquifer and stream drawdown derivatives and their dependence on $C_{D,r}$ across the range of values indicated. The graphs show the derivative plots of the NPS solutions in the pumped half-space at $(x_D, y_D) = (0.5, 0.0)$ and in the stream at $(-0.5W_D, 0.0)$. For the limiting case of $C_{D,r} \to \infty$, the derivative behavior shown is characteristic of models with fixed-head boundaries (Ferroud et al., 2019). For finite values of $C_{D,r}$, the derivative is characterized by two peaks associated with transition from early- to intermediate-time behavior and intermediate- to late-time behavior. The second peak is subsequently followed by a late-time steady decline to a constant value. Stream drawdown derivatives are characterized by a single unique peak for each value $C_{D,r}$.

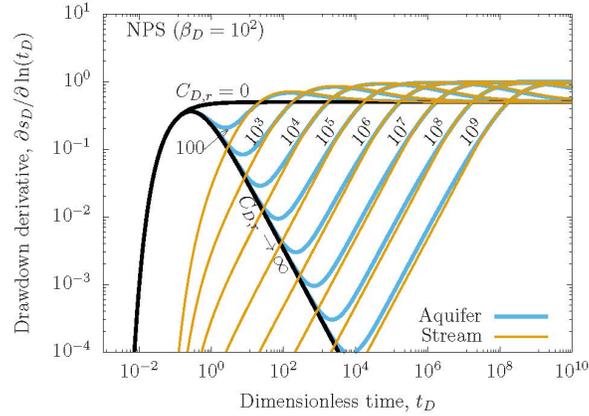

Figure 7. Plots of the temporal behavior of the log-time derivatives of aquifer and stream drawdown for different values of $C_{D,r}$. The plots are for the NPS solution with $\beta_D = 100$. The FPS solutions show a similar pattern.

The temporal behavior of the stream depletion rate (SDR) is shown in Figure 8, where the effect of the parameter $C_{D,r}$ on $Q_{D,r}$ is explored for the NPS solution, with $\beta_D = 10$. The limiting case of $C_{D,r} \to \infty$ is again included for comparison. For finite values of $C_{D,r}$, the stream depletion rate increases to a peak value before declining as pumping continues and drawdown of the stream increases. The maximum SDR decreases with decreasing values of $C_{D,r}$, and occur at earlier and earlier times. For large values of $C_{D,r}$ the stream can sustain groundwater pumping for more prolonged periods.

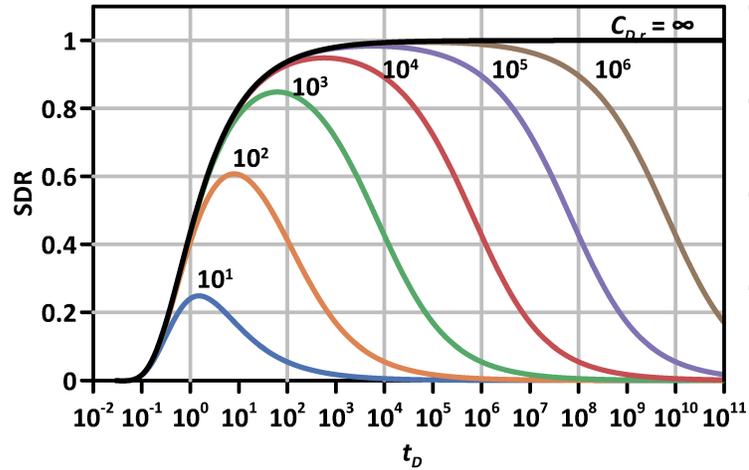

Figure 8. Temporal behavior of SDR obtained with the NPS model for different $C_{D,r}$ values in the range $C_{D,r} \in [10, \infty]$ with $\beta_D = 10$.

## Observed Transient Aquifer and Stream Drawdown

Aquifer drawdown obtained data by detrending and denoising observed water levels are shown in Figure 9 for the six pumping tests highlighted in Figure 3. All the aquifer drawdown data shown in Figure 9 are from the aquifer observation well and differ only by the time during the irrigation season at which the pumping occurred. They are plotted on log-log scale and for completeness, include the recovery period. Only drawdown data from the initial 24-hour period of each of the tests were analyzed with the NPS model.

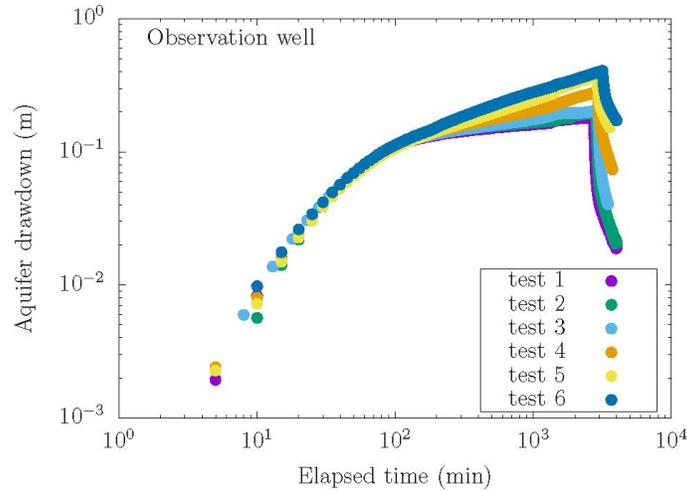

Figure 9. Log-log plot of aquifer drawdown recorded in the observation well during the six pumping tests reported in this work.

Observed stream stage drawdown data are shown in Figure 10 for the four monitoring stations along Stenner Creek for which data was available. The response of the aquifer in test 6 is also included in all the graphs for comparison. The data are plotted on log-log scale for (a) Stenner P1, (b) Stenner P2, (c) Stenner P3, and (d) Stenner P5, with elapsed time since onset of pumping on the horizontal axis and drawdown on the vertical axis. Recovery phase data are also included in the Figure for completeness. Only pumping-phase drawdown data, however, are considered in the analysis described in the following section. All observed stream drawdown responses are delayed relative to aquifer drawdown response even though the aquifer observation well is farther than all the stream monitoring locations. Additionally, all observed stream drawdown values are less in magnitude than the observed aquifer drawdown response. This is true even for location P1 that showed the largest stream drawdown response.

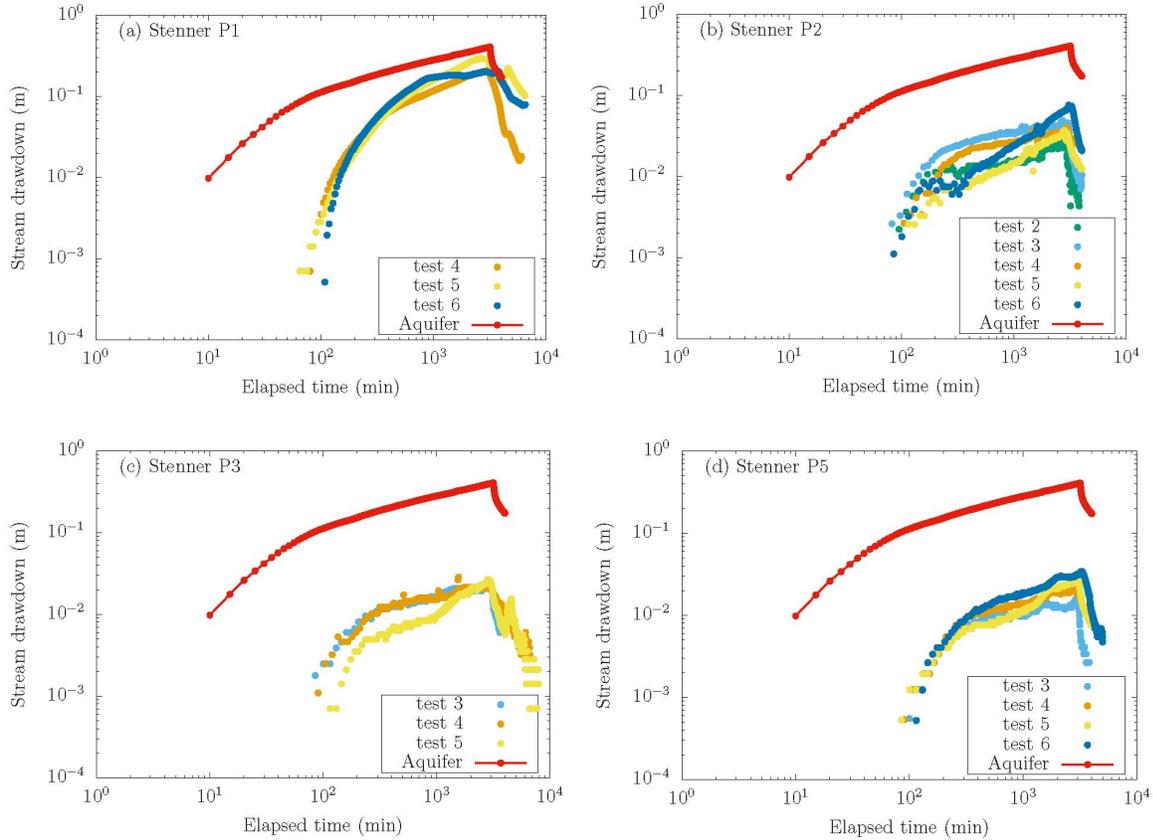

Figure 10. Log-log plots of transient stream stage drawdown response to pumping observed in stream channel stilling wells (a) Stenner-P1, (b) Stenner-P2, (c) Stenner-P3, and (d) Stenner-P5, for the respective tests recorded. Aquifer observation well data are included for comparison.

## Analysis of Aquifer and Stream Drawdown

The NPS model was used to analyze aquifer drawdown data shown in Figure 9. Only the parameters $K_x$, $S_s$, $\beta$, and $C_r$ were estimated. To simplify the analysis, horizontal anisotropy was fixed at $\kappa = 1.0$ for the results reported here. The results of the parameter estimation exercise are summarized in Table 3. The respective standard errors for each parameter estimate are included in the Table to quantify parameter estimation uncertainty. The corresponding NPS model fits to the drawdown data are shown in Figure 11, where they are plotted on (a) log-log and (b) semi-log scales.

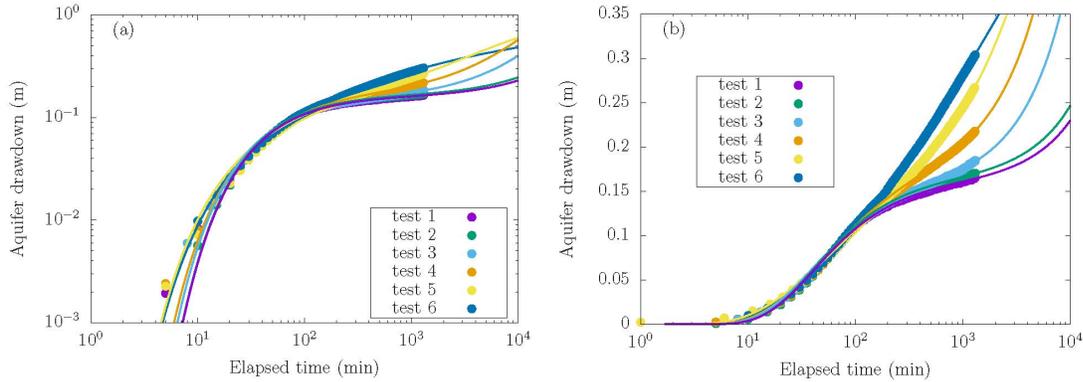

Figure 11. Results of transient analysis of aquifer drawdown response from individual pumping tests. The results show model fits to observation well (aquifer) drawdown data in (a) log-log scale and (b) semi-log scale. The dots are data and the solid lines are model fits to data.

Table 3. Estimated parameter values using aquifer drawdown data from the observation well, with fixed aquifer thickness $b = 11$ m, anisotropy ratio $\kappa = 1.0$, stream width $W = 1.5$ m, distance to pumping well $R = 60$ m, and distance to observation well $x = -15$ m.

| Test | $K_x$ ($\times 10^{-5}$ m/s) | $S_s$ ($\times 10^{-6}$ m$^{-1}$) | $\beta$ ($\times 10^{-5}$ s$^{-1}$) | $C_r$ (-) |
|---|---|---|---|---|
| 1 | $2.35 \pm 1.3$ | $4.02 \pm 1.9$ | $2.34 \pm 0.62$ | $86.8 \pm 49.4$ |
| 2 | $1.66 \pm 0.89$ | $3.22 \pm 1.5$ | $1.93 \pm 0.51$ | $79.9 \pm 39.5$ |
| 3 | $2.05 \pm 1.1$ | $3.86 \pm 1.7$ | $2.15 \pm 0.54$ | $22.6 \pm 5.66$ |
| 4 | $0.985 \pm 0.18$ | $2.42 \pm 0.39$ | $1.43 \pm 0.14$ | $8.32 \pm 0.64$ |
| 5 | $2.02 \pm 1.3$ | $29.1 \pm 1.2$ | $4.01 \pm 0.01$ | $6.43 \pm 0.15$ |
| 6 | 63.6 ± 5.5 | 55.2 ± 3.1 | 1.15 ± 0.35 | 0.70 ± 0.36 |

The parameter values in Table 3 obtained from analysis of aquifer drawdown were used to predict the expected stream drawdown behavior at Stenner-P1 and the results are shown in Figure 12. Given the model-data misfit, the NPS model was used to separately analyze stream drawdown data collected at Stenner-P1, giving estimates of system hydraulic parameters. Transient stream drawdown data were available for tests 4, 5 and 6. For this analysis, the aquifer anisotropy ratio and streambed conductance fixed at $\kappa = 1.0$ and $\beta = 10^{-5}$ s$^{-1}$, respectively, to limit model over-parametrization. The value of $\beta$ used here was from the analysis of aquifer drawdown data from test 6. The resulting model fits to stream drawdown data are shown in Figure 13. The estimated parameter values for $K_x$, $S_s$, and $C_r$ are summarized in Table 4.

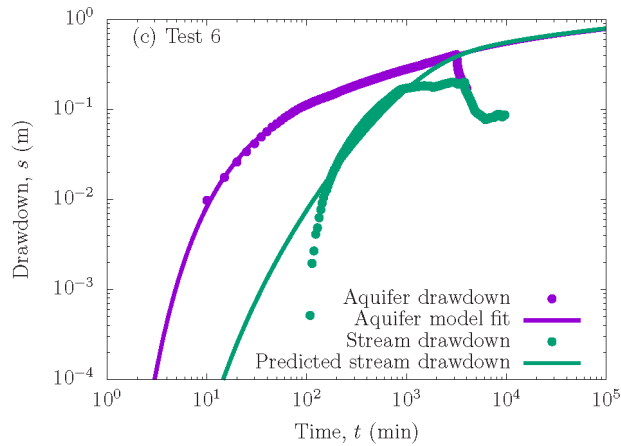

Figure 12. Transient stream drawdown predicted by the model using parameter values estimated from aquifer drawdown data for test 6. The corresponding model fit to aquifer data are included.

Table 4. Estimated parameter values using the NPS solution for the three tests for which data were collected at the Stenner-P1 stream stage monitoring location. Parameters $\kappa = 1.0$ and $\beta = 1.15 \times 10^{-5}$ s$^{-1}$ were fixed to values estimated with aquifer drawdown data from test 6. The standard error, $\pm\sigma_\epsilon$, for each parameter estimate is included to provide a measure of parameter estimation uncertainty.

| Test | $K_x$ ($\times 10^{-4}$ m/s) | $S_s$ ($\times 10^{-4}$ m$^{-1}$) | $C_r$ ($\times 10^{-2}$) |
|---|---|---|---|
| 4 | 8.81 ± 0.11 | 5.85 ± 0.28 | 2.56 ± 0.26 |
| 5 | 4.86 ± 0.02 | 5.23 ± 0.02 | 4.28 ± 0.11 |
| 6 | 3.93 ± 0.12 | 4.84 ± 0.09 | 2.35 ± 0.47 |

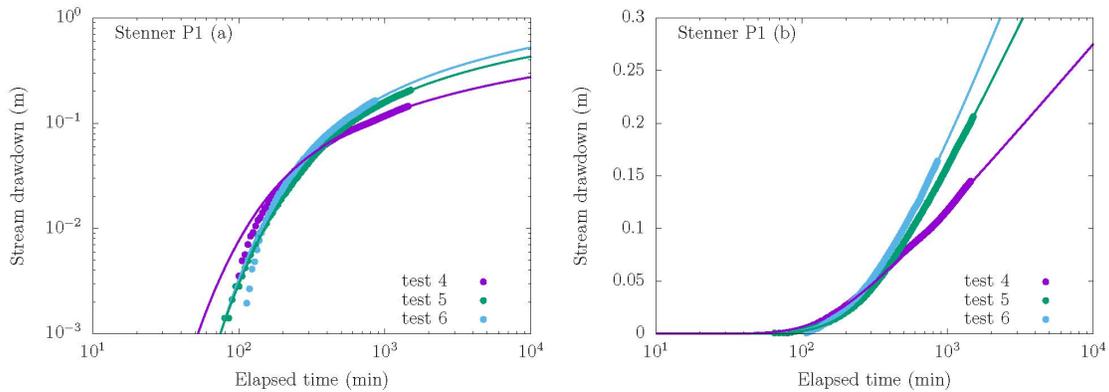

Figure 13. Results of transient analysis of stream drawdown response from individual tests showing model fits to the data for Stenner Creek stilling well S1. The results are plotted on (a) log-log and (b) semi-log scales. Dots are data and solid lines are model fits.

# Discussion

**Model Predicted Behavior**

The predicted aquifer drawdown response shown in Figure 4 is characterized by three phases that are commonly observed in media with two mechanisms of water storage such as unconfined aquifers (Boulton, 1963; Neuman, 1974; Malama, 2011; Lin et al., 2019) and dual-porosity fractured formations (Warren and Root, 1963; Streltsova, 1983; De Smedt, 2011; Lin and Yeh, 2021). Hunt (2009) referred to the behavior as a type of delayed yield drawdown in a pumped confined aquifer near a stream. The behavior has been observed by other workers including Fox et al. (2002), Hunt et al. (2001), Kollet and Zlotnik (2003), Lough and Hunt (2006), and Nyholm et al. (2002). During early-time, aquifer response follows the limiting models of Theis (1935) and Ferris et al. (1962) for the NPS and FPS models, respectively, which corresponding to $C_{D,r} = 0$. The onset of intermediate time is marked by significant departure of aquifer drawdown from limiting case of $C_{D,r} = 0$ transitioning to closely follow the behavior predicted by the limiting solutions of Fox et al. (2002) and Hantush (1965) for $C_{D,r} \to \infty$. During this transition period, stream appears to initially serve as a near-infinite store of water and aquifer drawdown appears to approach a "steady state." For finite values of $C_{D,r}$, this "steady state" phase is only momentary as aquifer drawdown begins to increase again and stream stage begins to respond to pumping yielding measurable transient stream drawdown. If pumping were to cease during the early part of intermediate-time phase, the models of Fox et al. (2002) and Hantush (1965) would be sufficient to describe system behavior. The higher the value of $C_{D,r}$, the longer this quasi-static phase persists as leakage from stream storage offsetting some of the drawdown attributable to aquifer storage depletion. As pumping continues, however, aquifer drawdown response transitions into late-time transient behavior characterized by increasing aquifer drawdown above the steady-state values predicted by the fixed-stage models of Fox et al. (2002) and Hantush (1965), which shows that these models underestimate aquifer drawdown at late-time. At very late-time, aquifer drawdown transitions into the behavior predicted by models with a no-flow condition at the stream-aquifer interface. This late-time behavior is a return to water withdrawal from aquifer elastic storage if the stream is completely drawn down.

Model predicted results also indicate that stream stage drawdown response is delayed relative to aquifer drawdown, and only appears to start during the late-time phase. However, stream stage responsiveness to pumping increases with increasing values of $\beta_D$, leading to decreasing lag times between stream and aquifer drawdown responses. Stream drawdown is further shown to exceed the late-time steady-state aquifer drawdown predicted by the fixed-stage models of Fox et al. (2002) and Hantush (1965). Depending on the initial stage of the stream, stream drawdown would lead to drying up of streams at very late time, during prolonged periods of groundwater pumping. Stream drawdown is accompanied by a depletion rate that increases initially with time before reaching a peak rate followed by a subsequent decline. In this instance, the solution of Zlotnik (2004) serves as the limiting case of depletion rate. The behavior of $Q_{D,r}$ for finite values of $C_{D,r}$ differs significantly from that predicted for $C_{D,r} \to \infty$ where the maximum depletion rate stays fixed at $Q_{D,r} = 1.0$ indefinitely, with all water captured by the pumping well at late time coming from stream recharge.

**Drawdown Data Analysis**

Analysis of drawdown data from the observation well shows a general shift in aquifer drawdown behavior over the course of the six pumping events during the summer 2022 irrigation season. The least overall drawdowns were recorded in the first of the pumping periods of March 16 to March 18, while the largest were recorded in the last reported pumping period. This behavior is reflective of the decrease in the amount of water stored in the stream channel ($C_r$) from the spring high flows to the summer lows. However, for any given test, $C_r$ may be treated as constant. The observed aquifer drawdown generally shows the three phases predicted by the model developed in this work and does not attain late-time steady state predicted by the fixed-stage models of Hantush (1965) and Fox et al. (2002).

Another important observation of note is that a nonzero, coherent, and unambiguous transient stream drawdown response is clear in the data. For our study site, the stream clearly does not behave as having a fixed stage and has a finite capacity to supply recharge during pumping. Additionally, the observed stream drawdown response at all measuring locations is significantly delayed relative to aquifer drawdown. All the stream stage observation locations are located closer to the pumping well than the aquifer observation well but all start showing measurable drawdown response at much later times (over an hour later) than the aquifer observation well. This delayed stream response confirms a key model prediction shown in Figure 4.

Results of the parameter estimation exercise based on aquifer drawdown data demonstrate that excellent model fits to data with $R^2 \geq 95\%$ and estimated parameter values are reasonable for a sand and gravel aquifer. The exercise also shows that reasonable streambed conductance and stream channel storage values are estimable from data using the model developed in this work. The parameter estimation uncertainties decreased over the six tests as the three phase of aquifer drawdown response became more defined as stream channel storage decreased.

The transient stream drawdown behavior predicted based on hydraulic parameters estimated from only from aquifer drawdown data deviates significantly from the observed behavior at Stenner-P1. This suggests the need for joint analysis of aquifer and stream drawdown data, or for separate analysis of stream drawdown data. In this work, stream drawdown data were analyzed separately and the resulting model fits to stream drawdown data showed marked improvement. The results also suggest that local aquifer and streambed hydraulic properties in the vicinity of the observation location have a strong effect on the estimated model parameters. The results also highlight a deficiency in the theory developed here where stream flow velocity is neglected in the stream channel mass balance condition.

## Conclusions

It is demonstrated in this work that streams can have finite and estimable channel storage and that they can undergo both depletion and measurable drawdown in response to groundwater pumping from a hydraulically connected aquifer. This is especially critical in aquifer systems subjected to prolonged groundwater abstraction, which can lead to the drying of streambeds, as has been observed in many groundwater basins with irrigated agriculture. Models with fixed stream stage overestimate the available groundwater supply from the stream because of their inherent assumption of infinite stream storage. The results of this work have implications for sustainable groundwater management. The model developed may be used to not only predict the stream depletion rate but also the decline of stream stage in response to groundwater pumping. Additional work is needed to incorporate routing and stream discharge (or velocity) in the model and to conduct longer pumping tests than reported herein to better constrain parameter estimates. This may resolve the discrepancy in model parameter values estimated from independent analyses of aquifer and stream drawdown data. Another limitation of the models developed in this work is the assumption that vertical flow is negligible, which may not apply to aquifers with leakage across the confining layers where vertical flow can play a significant role.

## Acknowledgements

This paper describes objective technical results and analysis. Any subjective views or opinions that might be expressed in the paper do not necessarily represent the views of the U.S. Department of Energy or the United States Government.

This article has been co-authored by an employee of National Technology & Engineering Solutions of Sandia, LLC under Contract No. DE-NA0003525 with the U.S. Department of Energy (DOE). The employee owns all rights, title and interest in and to the article and is solely responsible for its contents. The United States Government retains and the publisher, by accepting the article for publication, acknowledges that the United States Government retains a non-exclusive, paid-up, irrevocable, world-wide license to publish or reproduce the published form of this article or allow others to do so, for United States Government purposes. The DOE will provide public access to these results of federally sponsored research in accordance with the DOE Public Access Plan https://www.energy.gov/downloads/doe-public-access-plan.

Ying-Fan Lin thanks the John Su Foundation for its support during this study.

## Code Availability

Mathematica and MATLAB scripts developed for the study are available at the hyperlinks: http://www.hydroshare.org/resource/e82cc60145e64b54bbe64dbdadd14d1d and https://www.mathworks.com/matlabcentral/fileexchange/118155-transient-stream-drawdown-and-depletion. The raw data analyzed in this work are available from the corresponding author upon request.

## Supporting Information

Additional Supporting Information may be found in an online document as referenced in the text above. Supporting Information is generally not peer reviewed.

# References

Asadi-Aghbolaghi, M. and Seyyedian, H. 2010. An analytical solution for groundwater flow to a vertical well in a triangle-shaped aquifer. *Journal of Hydrology* 393, no. 3–4: 341–348.

Boulton, N. 1963. Analysis of data from non-equilibrium pumping tests allowing for delayed yield from storage. *Proceedings of the Institution of Civil Engineers* 26, no. 3: 469–482.

Bourdet, D., Whittle, T.M., Douglas, A.A. and Pirard, Y.M. 1983. A new set of type curves simplifies well test analysis. *World Oil 196*, no. 6: 95-106.

Bowen, J.L., Kroeger, K.D., Tomasky, G., Pabich, W.J., Cole, M.L., Carmichael, R.H. and Valiela, I. 2007. A review of land–sea coupling by groundwater discharge of nitrogen to New England estuaries: mechanisms and effects. *Applied Geochemistry* 22, no. 1: 175-191.

Butler Jr, J.J., Zhan, X. and Zlotnik, V.A. 2007. Pumping-induced drawdown and stream depletion in a leaky aquifer system. *Groundwater* 45, no. 2: 178–186.

Butler Jr, J.J., Zlotnik, V.A. and Tsou, M.-S. 2001. Drawdown and stream depletion produced by pumping in the vicinity of a partially penetrating stream. *Groundwater* 39, no. 5: 651–659.

Chan, Y. 1976. Improved image-well technique for aquifer analysis. *Journal of Hydrology* 29, no. 1–2: 149–164.

Chow, V.T. 1952. On the determination of transmissibility and storage coefficients from pumping test data. *Eos, Transactions American Geophysical Union* 33, no. 3: 397–404.

Cooper Jr, H. and Jacob, C.E. 1946. A generalized graphical method for evaluating formation constants and summarizing well-field history. *Eos, Transactions American Geophysical Union* 27, no. 4: 526–534.

De Smedt, F. 2011. Analytical solution for constant-rate pumping test in fissured porous media with double-porosity behaviour. *Transport in Porous Media* 88, no. 3: 479–489.

Faouzi, J. and Janati, H. 2020. pyts: A Python Package for Time Series Classification. *Journal of Machine Learning Research* 21, no. 46: 1–6.

Ferris, J.G., Knowles, D.B. and Brown, R.H. 1962. *Theory of aquifer tests*. *U.S. Department of the Interior, U.S. Geological Survey.* Water-Supply Paper 1536-E.

Ferroud, A., Chesnaux, R. and Rafini, S. 2018. Insights on pumping well interpretation from flow dimension analysis: The learnings of a multi-context field database. *Journal of Hydrology* 556: 449–474.

Ferroud, A., Rafini, S. and Chesnaux, R. 2019. Using flow dimension sequences to interpret non-uniform aquifers with constant-rate pumping-tests: A review. *Journal of Hydrology X* 2: 100003.

Foglia, L., McNally, A. and Harter, T. 2013. Coupling a spatiotemporally distributed soil water budget with stream-depletion functions to inform stakeholder-driven management of groundwater-dependent ecosystems. *Water Resources Research* 49, no. 11: 7292–7310.

Fox, G.A., DuChateau, P. and Dumford, D.S. 2002. Analytical model for aquifer response incorporating distributed stream leakage. *Groundwater* 40, no. 4: 378–384.

Glover, R.E. and Balmer, G.G. 1954. River depletion resulting from pumping a well near a river. *Eos, Transactions American Geophysical Union* 35, no. 3: 468–470.

Haberman, R. 2012. *Applied partial differential equations with Fourier series and boundary value problems*. Pearson Higher Ed.

Hantush, M.S. 1965. Wells near streams with semipervious beds. *Journal of Geophysical Research* 70, no. 12: 2829–2838.

Harbaugh, A.W. 2005. MODFLOW-2005, the U.S. Geological Survey modular ground-water model: the ground-water flow process. *U.S. Department of the Interior, U.S. Geological Survey*, Techniques and Methods, 6-A16.

Huang, C.S., Wang, Z., Lin, Y.C., Yeh, H.D. and Yang T., 2020. New analytical models for flow induced by pumping in a stream-aquifer system: A new Robin boundary condition reflecting joint effect of streambed width and storage. *Water Resources Research 56*, no. 4: e2019WR026352.

Huang, C.-S., Yang, T. and Yeh, H.-D. 2018. Review of analytical models to stream depletion induced by pumping: Guide to model selection. *Journal of Hydrology* 561: 277–285.

Hunt, B. 2003. Field-data analysis for stream depletion. *Journal of Hydrologic Engineering* 8, no. 4: 222–225.

Hunt, B. 2008. Stream depletion for streams and aquifers with finite widths. *Journal of Hydrologic Engineering* 13, no. 2: 80–89.

Hunt, B. 2009. 'Stream depletion in a two-layer leaky aquifer system', *Journal of Hydrologic Engineering*, 14, 9: 895–903.

Hunt, B., Weir, J. and Clausen, B. 2001. A stream depletion field experiment. *Ground Water* 39, no. 2: 283.

Intaraprasong, T. and Zhan, H. 2009. A general framework of stream–aquifer interaction caused by variable stream stages. *Journal of Hydrology* 373, no. 1–2: 112–121.

Jenkins, C.T. 1968. Techniques for Computing Rate and Volume of Stream Depletion by Wells. *Ground Water* 6, no. 2: 37–46.

Kollet, S.J. and Zlotnik, V.A. 2003. Stream depletion predictions using pumping test data from a heterogeneous stream–aquifer system (a case study from the Great Plains, USA). *Journal of Hydrology* 281, no. 1–2: 96–114.

Kwon, H.I., Koh, D.C., Jung, Y.Y., Kim, D.H. and Ha, K. 2020. Evaluating the impacts of intense seasonal groundwater pumping on stream–aquifer interactions in agricultural riparian zones using a multi-parameter approach. *Journal of Hydrology* 584: 124683.

Langevin, C.D., Hughes, J.D., Banta, E.R., Niswonger, R.G., Panday, S. and Provost, A.M., 2017. Documentation for the MODFLOW 6 groundwater flow model. *U.S. Department of the Interior, U.S. Geological Survey* Techniques and Methods, 6, A55: 197.

Laszuk, D. 2017. Python implementation of Empirical Mode Decomposition algorithm. *GitHub Repository*. GitHub.

Lin, Y.-C., Huang, C.-S. and Yeh, H.-D. 2019. Analysis of unconfined flow induced by constant rate pumping based on the lagging theory. *Water Resources Research* 55, no. 5: 3925–3940.

Lin, Y.-C. and Yeh, H.-D. 2021. An Analytical Model with a Generalized Nonlinear Water Transfer Term for the Flow in Dual-Porosity Media Induced by Constant-Rate Pumping in a Leaky Fractured Aquifer. *Water Resources Research* 57, no. 8: e2020WR029186.

Lough, H.K. and Hunt, B. 2006. Pumping test evaluation of stream depletion parameters. *Groundwater* 44, no. 4: 540–546.

Malama, B. 2011. Alternative linearization of water table kinematic condition for unconfined aquifer pumping test modeling and its implications for specific yield estimates. *Journal of Hydrology* 399, no. 3–4: 141–147.

Niswonger, R.G. and Prudic, D.E., 2005. Documentation of the streamflow-routing (SFR2) package to include unsaturated flow beneath streams--A modification to SFR1. *U.S. Department of the Interior, U.S. Geological Survey* Techniques and Methods 6, A13: 50.

Neuman, S.P. 1974. Effect of partial penetration on flow in unconfined aquifers considering delayed gravity response. *Water Resources Research* 10, no. 2: 303–312.

Nyholm, T., Christensen, S. and Rasmussen, K.R. 2002. Flow depletion in a small stream caused by ground water abstraction from wells. *Groundwater* 40, no. 4: 425–437.

Owen, D., Cantor, A., Nylen, N.G., Harter, T. and Kiparsky, M. 2019. California groundwater management, science-policy interfaces, and the legacies of artificial legal distinctions. *Environmental Research Letters* 14, no. 4: 045016.

Panday, S., Langevin, C.D., Niswonger, R.G., Ibaraki, M. and Hughes, J.D., 2013. MODFLOW–USG version 1: An unstructured grid version of MODFLOW for simulating groundwater flow and tightly coupled processes using a control volume finite-difference formulation. *Department of the Interior, U.S. Geological Survey* Techniques and Methods, 6-A45.

Prudic, D.E. 1989. Documentation of a computer program to simulate stream-aquifer relations using a modular, finite-difference, ground-water flow model. *Department of the Interior, U.S. Geological Survey* Open-File Report, 88-729.

Prudic, D.E., Konikow, L.F. and Banta, E.R. 2004. A new streamflow-routing (SFR1) package to simulate stream-aquifer interaction with MODFLOW-2000. *Department of the Interior, U.S. Geological Survey* Open-file report 2004-1042.

Refsgaard, J.C., Storm, B. and Clausen, T. 2010. Système Hydrologique Europeén (SHE): review and perspectives after 30 years development in distributed physically-based hydrological modelling. *Hydrology Research* 41, no. 5: 355.

Stehfest, H. 1970. Algorithm 368: Numerical inversion of Laplace transforms [D5]. *Communications of the ACM* 13, no. 1: 47–49.

Streltsova, T.D. 1983. Well pressure behavior of a naturally fractured reservoir. *Society of Petroleum Engineers Journal* 23, no. 05: 769–780.

Theis, C.V. 1935. The relation between the lowering of the piezometric surface and the rate and duration of discharge of a well using ground-water storage. *Eos, Transactions American Geophysical Union* 16, no. 2: 519–524.

Theis, C.V. 1941. The effect of a well on the flow of a nearby stream. *Eos, Transactions American Geophysical Union* 22, no. 3: 734–738.

Tolley, D., Foglia, L. and Harter, T. 2019. Sensitivity Analysis and Calibration of an Integrated Hydrologic Model in an Irrigated Agricultural Basin with a Groundwater-Dependent Ecosystem. *Water Resources Research* 55, no. 9: 7876–7901.

Warren, J. and Root, P.J. 1963. The behavior of naturally fractured reservoirs. *Society of Petroleum Engineers Journal* 3, no. 03: 245–255.

Winter, T.C., Harvey, J.W., Franke, O.L., and Alley, W.M. 1998. Ground water and surface water: a single resource. *Department of the Interior, U.S. Geological Survey* Circular 1139

Xiong, M., Tong, C. and Huang, C.-S. 2021. A New Approach to Three-Dimensional Flow in a Pumped Confined Aquifer Connected to a Shallow Stream: Near-Stream and Far-From-Stream Groundwater Extractions. *Water Resources Research* 57, no. 4: e2020WR028780.

Yu, H.-L. and Chu, H.-J. 2010. Understanding space–time patterns of groundwater system by empirical orthogonal functions: a case study in the Choshui River alluvial fan, Taiwan. *Journal of Hydrology* 381, no. 3–4: 239–247.

Zipper, S.C., Dallemagne, T., Gleeson, T., Boerman, T.C. and Hartmann, A. 2018. Groundwater pumping impacts on real stream networks: Testing the performance of simple management tools. *Water Resources Research* 54 no. 8: 5471-5486.

Zlotnik, V.A. 2004. A concept of maximum stream depletion rate for leaky aquifers in alluvial valleys. *Water Resources Research* 40, no. 6.

Zlotnik, V.A. and Huang, H. 1999. Effect of shallow penetration and streambed sediments on aquifer response to stream stage fluctuations (analytical model). *Ground Water* 37, no. 4: 599–599.

Zlotnik, V.A., Huang, H. and Butler Jr, J.J. 1999. Evaluation of stream depletion considering finite stream width, shallow penetration, and properties of streambed sediments. *2nd International Conference on Water Resources & Environment Research* [Preprint].

Zlotnik, V.A. and Tartakovsky, D.M. 2008. Stream depletion by groundwater pumping in leaky aquifers. *Journal of Hydrologic Engineering* 13, no. 2: 43–50.